\documentstyle[prd,aps]{revtex}
\input epsf.tex
\def\DESepsf(#1 width #2){\epsfxsize=#2 \epsfbox{#1}}
\begin{document}

\draft
\preprint{{}}
\title{Phenomenology of light remnant doubly charged Higgs fields\\ in
 the supersymmetric left-right model}
\author{B. Dutta$^1$ and R. N. Mohapatra$^2$ }
\address{$^1$Institute of Theoretical Sciences, University of Oregon, Eugene, 
OR-97403, USA.\\
$^2$ Department of Physics, University of Maryland, College Park, MD-20742, 
USA.}
\date{April, 1998}
\maketitle
\begin{abstract} It has recently shown that in supersymmetric left-right models
with automatic R-parity conservation, the theory below the $W_R$ scale is given
by MSSM with massive neutrinos and a pair of doubly charged superfields with
masses in the 100 GeV range ( with or without an extra pair of heavy Higgs
doublets  (M$\geq$ 10 TeV) depending on the model). In this paper we study the 
unification prospects  for such theories and their phenomenological 
implications for collider experiments. We study two versions of the theory, one
with supersymmetry  breaking transmitted via the gauge and another where the
same occurs via  gravitational forces. We point out that  looking at multi
$\tau$  final states can considerably constrain the parameter space of the
model. 
\end{abstract} \pacs{\hskip 6 cm  OITS-649;  UMD-PP-98-109 }

\vskip2pc

\section{Introduction}

Supersymmetric left-right models (SUSYLR) where the $SU(2)_R$ gauge symmetry is
broken by triplet Higgs fields with $B-L=2$ have many  attractive features: 1)
they imply automatic conservation of baryon and lepton  number\cite{moh}, a
property which makes the standard model so attractive,   but is not shared by 
the minimal supersymmetric standard model (MSSM);  2) they provide a natural 
solution to the strong and weak CP problems of the
   MSSM\cite{rasin}; 3) They yield a natural embedding of the see-saw mechanism
for small  neutrino masses\cite{gell} where the right handed triplet  field
($\Delta^c$) that breaks the $SU(2)_R$ symmetry also gives heavy mass to the
righthanded  Majorana neutrino needed for implementing the see-saw mechanism.
   
In order to cancel anomalies as well as to maintain supersymmetry below the
$SU(2)_R$ scale ($v_R$), one needs a pair of fields  $\Delta^c\oplus
\overline{\Delta^c}$ with $B-L=\mp 2$ respectively. An important   
distinguishing characteristic of these Higgs multiplets is that they contain
doubly charged Higgs  bosons and Higgsinos in them  that remain as physical
fields subsequent to symmetry breakdown. It has  recently been shown that the
vacuum of the   theory may or may not  conserve  R-parity\cite{kuchi}. If
however we require the   ground state to conserve R-parity , one must include
higher dimensional operators\cite{rasin,aulakh} or additional Higgs fields 
which break parity\cite{kuchi,goran2}. In this case,
\cite{goran2,chacko} the model in its simple versions, always predicts  that
some of the doubly charged fields mentioned above are massless  in the  absence
of  the higher dimensional operators (HDO). This is independent  of whether the
hidden sector supersymmetry breaking scale is above or  below the $W_R$ scale.
In the presence of HDO's, they  acquire masses of order 
$\sim v^2_R/M_{Pl}$.  Since  the measurement of the  Z-width at LEP and SLC
implies that such particles  must  have a mass of at least 45 GeV, this  puts a
lower limit on the $W_R$ scale of about
$10^{10}$ GeV or so. This result is interesting since $W_R$ masses  in this 
range also lead to neutrino masses expected on the basis of current solar and
atmospheric  neutrino experiments. If we take  the $W_R$ mass to be close to 
this lower limit (say of order $10^{10}-10^{11}$ GeV), it implies the masses of
the doubly charged particles in the 100 GeV range. There are two bosonic and two
fermionic particles of this type. The rest of the particle spectrum below the
$W_R$ scale can be same as that of the minimal supersymmetric standard model
(MSSM) with a massive neutrino or  it can have an extra pair of Higgs doublets 
in the 10 TeV  range depending on the structure of the model.  It is the goal of
this  paper to explore the constraints on the parameters of the model and
suggest tests in the $e^+e^-$ and $p\bar{p}$ collider.

Our main results are that for a large range of parameters in this model  the
stau (the superpartner of $\tau$)  is the lightest of the sleptons due to the
renormalization group running arising from the $\Delta^{++}\tau^-\tau^-$
coupling which is not very constrained from phenomenology. (Note that the
$\Delta^{++}$ couplings to other leptons are severely bounded by the recent PSI
results on muonium- antimuonium oscillation\cite{jungman}). As a result, tau
lepton final states, generated from the production of the
$\Delta^{\pm\pm}$ and its fermionic part, provide a crucial signature of this
class of models. For instance we find that detection of final states of type
$\tau^-\tau^-\tau^+\tau^+\gamma
\gamma$ plus missing energy or $\tau^-\tau^-\tau^+\tau^+$ with or without 
missing energy in both 
$p\bar{p}$ and $e^+e^-$ collsion will provide test of these models. Thus
non-observation of such signals will significantly reduce the domain of 
allowed   
parameters of this model. We  point out the difference between the multi $\tau$ 
signals of this model and the same type of signals that appear in the other 
models e. g. conventional GMSB type. We will show that the allowed region of
parameter space is larger when we choose GMSB type of theories.  We also
discuss the unification  prospects of this model in different SUSY breaking
scenarios.

This paper is organized as follows: in section 2, we review the  arguments
leading to the existence of  light doubly charged fields despite a high $W_R$
scale in the context of a simple model and  discuss the low energy interactions
of these fields; in  section 3 we discuss gauge unification in these  
models; in section  4, We discuss its parameter space and experimental 
signals in gravity mediated SUSY breaking scenario; in section 5, we 
discuss the same for the gauge mediated SUSY breaking scenarios and
in section 6 we present our conclusion.          

\section{Overview of the model }

In this section, we present a brief review of the arguments leading to the
existence of the light doubly charged  Higgs fields in the SUSYLR model. In 
order to give our arguments, we start by giving the basic features of  the
model, which is based on the gauge group
$SU(2)_L\times SU(2)_R\times U(1)_{B-L}\times SU(3)_c$. In Table I, we give the
particle content of the model. We will  suppress the $SU(3)_c$ indices in what
follows. \\\vskip2mm

\begin{tabular}{|c|c|c|} \hline Fields  & SU$(2)_L \, \times$ SU$(2)_R
\, \times$ U$(1)_{B-L}$ &  group transformation \\
                 & representation & \\ \hline Q & (2,1,$+ {1
\over 3}$) & UQ \\
$Q^c$  & (1,2,$- {1 \over 3}$) & VQ$^c$ \\ L  & (2,1,$- 1$) & UL \\
$L^c$ & (1,2,+ 1) & V$L^c$ \\
$\Phi_{1,2}$ & (2,2,0) & $U\phi V^{\dagger}$ \\
$\Delta$& (3,1,+ 2) & U$\Delta U^{\dagger}$ \\
$\bar{\Delta}$ & (3,1,$- 2$) & $U\bar{\Delta}U^{\dagger}$ \\
$\Delta^c$ & (1,3,+ 2) & $V\Delta^c V^{\dagger}$ \\
$\bar{\Delta}^c$ & (1,3,$- 2$) & $ V\bar{\Delta}^cV^{\dagger}$ \\ 
$S$&  (1,1,0) &  $S$ \\\hline
\end{tabular}\\ Table 1:Field content of the SUSY LR model; we assume that $S$
\\is odd  under parity; $U$ and $V$ denote the $SU(2)_{L,R}$ transformations 
respectively. 
\vskip 4mm

The superpotential for this theory is given by (we have suppressed the
generation index):
                 
\begin{eqnarray}  W & = & {\bf h}^{(i)}_q Q^T \tau_2 \Phi_i \tau_2 Q^c + {\bf
h}^{(i)}_l L^T \tau_2 \Phi_i \tau_2 L^c
\nonumber\\
  & +  & i ( {\bf f} L^T \tau_2 \Delta L + {\bf f}_c {L^c}^T \tau_2 \Delta^c 
L^c)
\nonumber\\
  & +  & M_{\Delta} [{\rm Tr} ( \Delta \bar{\Delta} ) +
 {\rm Tr} ( \Delta^c \bar{\Delta}^c )] +\lambda S(\Delta\overline{\Delta}
-\Delta^c\overline{\Delta^c}) + \mu_S S^2 \nonumber\\
 & + & 
\mu_{ij} {\rm Tr} ( \tau_2 \Phi^T_i \tau_2 \Phi_j )+ W_{\it NR}
\label{eq:superpot}
\end{eqnarray}  where $W_{\it NR}$ denotes non-renormalizable terms arising from
higher scale physics such as grand unified theories or Planck scale effects.

\begin{eqnarray}
   W_{\it NR}= A [Tr(\Delta^c \overline {\Delta}^c)]^2/2  + B Tr(\Delta^c 
\Delta^c) Tr (\overline{\Delta}^c \overline{\Delta}^c)/2
\end{eqnarray}  where A and B are of order 1/$M_{Planck}$.

We will work in the vacuum which conserves R-parity. The Higgs vevs then have
the following pattern:
\begin{eqnarray} <\phi>=\left(\begin{array}{cc}
\kappa & 0\\ 0 & \kappa '\end{array}\right); <\Delta^c>=\left(\begin{array}{cc}
0 & v \\ 0 & 0 \end{array}\right)
\end{eqnarray} Similar pattern for $<\overline{\Delta^c}>$ is assumed.

Using Eq. 1, one can give a group theoretical argument for the existence of 
light doubly charged particles in the supersymmetric limit as follows. Let us
first ignore the higher dimensional terms $A$ and $B$ as well as the leptonic 
couplings $f$. It is then clear that the superpotential has a  complexified
$U(3)$ symmetry (i.e. a $U(3)$ symmetry whose parameters are taken to be
complex) that operates on the $\Delta^c$ and $\bar\Delta^c$ fields. This is due
to the holomorphy of the superpotential. After one  component of each of the
above fields acquires vev the resulting symmetry is the complexified 
$U(2)$. This leaves 10 massless fields. Once we bring in the D-terms and switch
on the gauge fields, six of these fields become massive as a consequence of the
Higgs mechanism of supersymmetric theories. That leaves four massless fields in
the absence of higher dimensional  terms. These are the two complex doubly
charged fields. Of the two non-renormalizable terms $A$ and $B$, only the A-term
has  the complexified $U(3)$ symmetry. Hence the  supersymmetric contribution to
the doubly charged particles will come only from the B-term. It is then clear
that the masses of the doubly charged fields are of order $v^2_R/M_{Pl}$.
Requiring that these masses satisfy the Z-width bound then implies that $v_R\geq
10^{10}-10^{11}$ GeV or so. In this paper we will assume that $v_R$ is at the
lower bound value so that the doubly charged fields are accessibly to the
existing collider  experiments. Note incidentally that although the leptonic
couplings do not  respect the above mentioned   symmetry, they are unimportant 
in determining the vacuum structure as  long as R-parity is conserved and hence
they do not effect the doubly  charged field masses.

Let us now give an explicit calculation of the masses of the doubly charged 
fields in the supersymmetric limit using the superpotential in Eq.(1).  Let us
write down the F-terms for the $S$, $\Delta$ and $\Delta^c$ terms:
\begin{eqnarray}  F_S=2\mu_S S +\lambda (\Delta
\overline{\Delta}-\Delta^c\overline{\Delta^c})
\nonumber \\ F_{\Delta} = (\lambda S + M_{\Delta} ) \overline{\Delta} \nonumber
\\ F_{\Delta^c}= (-\lambda S + M_{\Delta})\overline{\Delta^c} 
\end{eqnarray}  If the effective supersymmetry breaking scale is below the $W_R$
scale, then these $F$ terms must vanish. It is then clear that if we choose the
$\Delta^c$ and 
$\overline{\Delta^c}$ vev's (denoted by $v_R$ and $\bar{v}_R$) to be 
nonvanishing (and they are equal in the supersymmetric limit), then we must 
have $<S>=M_{\Delta}/\lambda$. This implies that the $\Delta$ and
$\overline{\Delta}$ vev's vanish and the masses of these fields are of order
$2M_{\Delta}$. Thus the left triplet fields decouple from the  low 
energy spectrum. It is then easy to see from the superpotential (in the absence
of the A and B terms) that all the particles in the superfields $\Delta^c$ and
$\overline{\Delta^c}$ are massless in the limit of exact supersymmetry. 
One linear combination of the neutral fields and another ofsingly charged 
fields  disappear due to the Higgs mechanism. The remaining singly charged
and neutral Higgs fields pick up mass of order of $v_R$ and disappear
from the low energy spectrum.

 The story of the doubly charged fields is however
very different in this theory as has been shown in Ref.\cite{chacko}. Once
supersymmetry breaking is turned on but the higher dimensional terms $A$ and
$B$ are excluded   from  the analysis, these fields  acquire negative
mass-squares signalling the breakdown of electric charge. This problem is cured
as soon as the $A$ and $B$ terms are included. The doubly charged fields (all
of them) then acquire masses of order $v^2_R/M_{Pl}$ and the vacuum becomes
charge conserving.

Let us now discuss the Higgs doublet spectrum of the model at low energies.  
At the
$W_R$ scale, one generally takes two bi-doublet  fields $\phi$'s to make the 
model realistic.
In order  to get the MSSM at low energies, one must decouple one pair of
$H_{u}$ and $H_d$ from the low energy spectrum. This has been called
doublet-doublet splitting problem in literature. In the model without HDO
contributions, it is clear from the superpotential in Eq. (1) that doublet
Higgsino matrix is symmetric:
\begin{eqnarray} M_H= \left(\begin{array}{cc}
\mu_{11} & \mu_{12}\\
\mu_{12} & \mu_{22} \end{array}\right)
\end{eqnarray}  If we now do fine tuning to get one pair of $H_{u,d}$ at low
energies, the $H_{u,d}$ appear as identical combinations of the doublets in
$\phi_i$'s. As a result, at the MSSM level, we have proportionality of the 
$M_u$ and $M_d$ leading to vanishing CKM angles. This result holds even if we 
increase the number of bi-doublets arbitrarily and uses only the fact that
bilinear mass matrix $\mu_{ij}$ is symmetric. This in fact raises the
interesting possibility\cite{rabi} that all mixing angles in the quark and 
lepton sector may arise purely out of radiative corrections involving the  soft
breaking terms\cite{raby}. An advantage of this version of the model is that
there are no new flavor changing effects other than those from the usual
supersymmetric sources\cite{masiero}.

On the other hand, one may choose the $\mu_{ij}$ parameters of this model to be
of the order of electroweak scale so that the low energy model is not  exactly
the MSSM but rather the two Higgs pair extension of MSSM. The  phenomenology of
these models are very similar to the previous case except that there are new
contributions to the flavor changing neutral current effects in this model
similar to those in the nonsupersymmetric  left-right models\cite{grimus} which
puts a lower limit on the masses of the second pair of Higgs doublets to be in
the 5-10 TeV range. As a result they will essentially decouple from the low
energy spectrum. 

Our results are independent of which of the above choices for the Higgs  sector
is made, except that unification discussion applies only to the second version.

\section{Gauge unification } 

The presence of the doubly charged fields at low energies distinguishes the 
gauge 
coupling evolution in this model from the MSSM and one might expect  that one
will lose the unification property. It however turns out that the  gauge
couplings do unify in this model, albeit at a lower scale as we see  below,
for the case which has two pairs of Higgs doublets at the weak scale. The 
couplings evolve according to  their respective beta functions. As just 
mentioned, below the $v_R$ scale, we assume four Higgs
doublet fields instead of the usual two of MSSM and we assume that one 
of  the doublet pairs has mass of 10 TeV. They lead to a trivial
modification of the beta function below $v_R$. The beta functions above the
$v_R$ scale are given below for one loop :
\begin{eqnarray}
 b^{1223}_{i}=\left( \matrix{ 0\cr-6\cr -6\cr -9\cr} \right)  +N_F
\left( \matrix{  2\cr 2\cr 2\cr 2\cr} \right)  +n_{\Phi} \left( \matrix{ 0\cr 
1\cr 1\cr 0\cr} \right) +n_{\Delta} \left( \matrix{  9\cr 4\cr 0\cr 0\cr}
\right) + n_{\Delta^c}
 \left(
\matrix{  9\cr 0\cr 4\cr 0\cr} \right)  \ \ \, ,
\end{eqnarray} and in the following equation for the two-loop:
\begin{eqnarray} b^{1223}_{ij}&=&\left( \matrix{ 0&0&0&0\cr 0&-24&0&0\cr
0&0&-24&0\cr 0&0&0&-54\cr} \right)   + N_F \left( \matrix{ 7/3&3&3&8/3\cr 
1&14&0&8\cr 1&0&14&8\cr 1/3&3&3&68/3\cr}\right) + n_{\Phi} \left( \matrix{ 
0&0&{0}&0\cr 0&{7}&{3}&0\cr {0}&{3}&{7}&0\cr 0&0&0&0\cr}\right)+ n_{\Delta}
\left( \matrix{  54&72&0&{0}\cr 24&{48}& {0}&0\cr {0}&{0}&{0}&0\cr 0&0&0&0 \cr}
\right)\\\nonumber &&+n_{\Delta^c} \left( \matrix{  54&0& {72}&0\cr
0&{0}&{0}&0\cr {24}&{0}&{48}&0\cr 0&0&0&0\cr} \right)  \, , 
\end{eqnarray}  where $i=U(1)_{B-L},SU(2)_L,SU(2)_R,SU(3)_C$  respectively  in 
the matrices,
$N_F$ is the number of fermion generations. $N_F=3$ always and $n_{\phi}$  is 
the number of  bidoublets which we take to be 2. We also take
$n_\Delta$(${\Delta}
 + \bar\Delta$) and $n_{\Delta^c}$($\Delta^c+\bar\Delta^c$) to be 1.  Since the
$\Delta  ^{\pm\pm}$ leaks down  to the weak scale, we need to include its
contribution to the running of the  gauge couplings in between the weak and the
intermediate scale. 
 Since this field $\Delta$  has only hypercharge quantum number under 
 the SM representation, the hypercharge gauge coupling RGE gets an extra term 
of 24/5 in one loop and in the two loop, the hypercharge squared elements  
gets and additional factor 
$ 72\times 16/25$. 

We see from figure 1a.  that  the gauge couplings 
unify  at a scale $\sim 10^{11}$ GeV and the intermediate scale is 
$\sim 10^8$ GeV. We take $\alpha_c=0.118$, $\alpha=1/128.7$ and
$\sin^2\theta_w=0.2321$ at the weak scale. The low unification scale implies
that the proton decay constraint would rule out groups like SO(10). However the 
final unifying group can be $SU(3)\times SU(3)\times SU(3)$ or any group 
that conserves baryon number.   

In the
supergravity  motivated models, the gauge unification is necessary in order to
have gaugino mass unification. The masses of all the sparticles can then be
determined in terms of the parameters  e.g. universal scalar mass $m_0$ at the
unification, universal gaugino mass $m_{1/2}$ and the  triliniear
coefficient in the potential A's. We will assume the A's to be 0. 
We choose to assume the universality of $m_0$ at the unification scale
rather than at the Planck scale for two reasons: first is that we do
not know the theory above the unification scale and the nature of the
evolution of the parameters is obviously dependent on those details.
The second reason is that the experimental signature we are interested
in involves only the tau lepton and its partner and not the other super
particles of the theory. So even if we assumed universality of scalar 
masses at the Planck scale which would necessarily imply some
splitting between generations due to running between the Planck 
scale to the unification scale, our final conclusions will be uneffected
by this.

If we use 
GMSB (gauge mediated supersymmetry breaking models) models where SUSY breaking
is communicated to the observable sector by gauge mediation, the
soft susy breaking scalar and the gaugino masses  are generated at a scale
$\sim 10^{5}$ GeV by gauge interactions. In this type of models gauge
unification is not necessary in order to have unified gaugino  mass, since 
they are generated in one loop. In our calculation for the GMSB cases we do not
consider any gauge unification. However in figure 1b we show one example where
we have the gauge unification with a messenger sector composed of: one copy of 
$D_R+\bar D_R$, $U_R+\bar U_R$  and $L_L+\bar L_L$. The  couplings unify at
$10^{10.5}$ GeV, the intermediate scale is at $\sim 10^{5.2}$ GeV and the
$\alpha_R/\alpha_{B-L}$ is 1/2 at the intermediate scale.
\begin{figure}[htb]
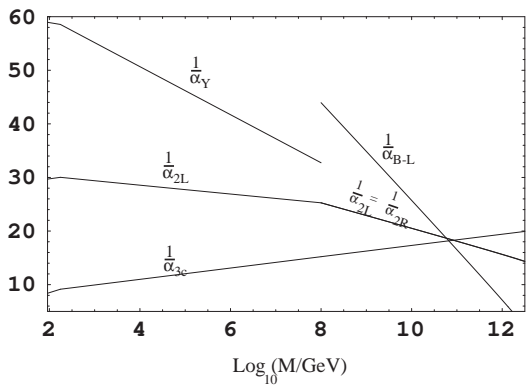

\vspace{0 cm}
\centerline{ \DESepsf(dmdeltafig1.epsf width 7 cm) }
\smallskip
\caption {Gauge coupling unification (two loop) in the Supergravity case is 
shown}
\end{figure}
\vspace{0 cm} 
\begin{figure}[htb]
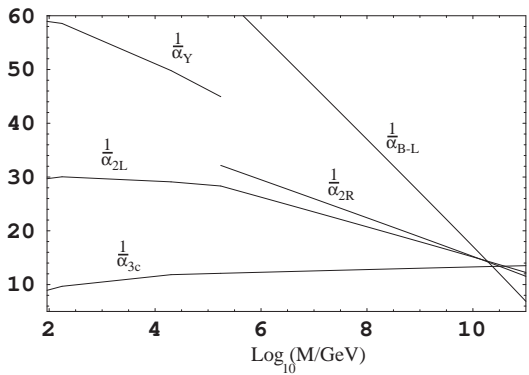

\vspace{1 cm}
\centerline{ \DESepsf(dmdeltafig2.epsf width 7 cm) }
\smallskip
\caption {Gauge coupling unification (two loop) in the GMSB case is shown}
\end{figure}          
\section{Gravity mediated scenario}

In the standard gravity mediated scenarios, one starts with a universal
mass-square for all scalar components of the chiral superfields at the Planck
scale and they are then extrapolated to the weak scale to determine their
masses\cite{martin}. For those fields with large Yukawa couplings  such as the
Higgs doublet $H_u$, top squark, the weak scale value is significantly lower
than the planck scale one. (In fact, for the $H_u$ turning negative gives  rise
to the celebrated phenomenon of radiative electroweak symmetry breaking). For
other squarks, the gaugino mass  contribution has the  effect of increasing 
their value  over the Planck  scale value. On the other hand for the sleptons,
the change between the  weak scale and Planck scale value is not very
significant since they  neither have large Yukawa couplings nor do they have
strong interactions.

Turning now to the SUSYLR model, as mentioned before, we will assume
universality at the unification and in contrast with the MSSM,
the effective theory below the right
handed scale, contains the following new coupling in the  superpotential:
\begin{eqnarray} W_{extra}=f_{i}{\Delta^c}^{\pm\pm}{\em l}_i{\em l}_i
\end{eqnarray}  
which will have a major impact on our spectrum at the weak scale. 

In writing the above coupling, we have used the fact 
that experimental limits on lepton flavor changing processes such as 
$\mu\rightarrow 3e$ and $\tau\rightarrow 3e$  imply that $f_{ij}$ with $i\neq j$
are very small compared to the diagonal  couplings $f_i$ and have therefore 
taken the liberty to simply drop the off-diagonal  couplings.  In what follows,
we  will denote $\Delta^c\equiv \Delta$. This interaction gives rise to the
process
$\mu^+e^-\rightarrow
\mu^-e^+$ with a strength $G_{M-\bar{M}}\simeq \frac{f_1 f_{2}}{4\sqrt{2}
M^2_{\Delta}}$. Recent PSI experiment\cite{jungman} has yielded an upper limit
on $G_{M-\bar{M}}\leq 3\times G_F\times 10^{-3}$. For
$M_{\Delta}=100 $ GeV, this implies that $f_1f_2\leq 1.2\times 10^{-3}$. Thus
we expect each of the couplings to be less than $.1$ barring  pathological
situations where only one of the couplings bears  the brunt of the constraint. 
On the other hand there is no such constraint on
$f_{3}$ from experiments. So we will choose it to be of order 0$.5$. 

The first implication of the relatively large $f_{3}$ is  on the weak  scale
value for the mass of the $\Delta$-boson. As shown in \cite{chacko}, at the
$v_R$ scale, both the $\Delta$ and its fermionic partner 
$\tilde{\Delta}$ ($\Delta$-ino) have nearly the same mass. The bosonic
component however runs  faster than the fermionic one. As a result, at low
energies we can expect that
$M_{\Delta} < M_{\tilde{\Delta}}$. This has important implications for 
phenomenology.  

A second consequence of the possible large $f_{3}$ is that the 
$\tilde{\tau^c}$  mass is drawn down to smaller  values at lower scales. In 
order to calculate the mass spectrum we use  the fermion masses, mixing
angles,  $f$'s and the magnitude of the gauge couplings as inputs at the weak
scale. We use
 $m_t=175$ GeV and $m_b=4.5 GeV$. We then use the RGE's  shown in the appendix
and in the reference \cite{barger} to determine the Yukawa and the gauge
couplings at the GUT scale. Then we use the universal boundary conditions and
use the RGE's for the soft SUSY breaking masses and the couplings to  determine
the masses at the weak scale. We determine  the parameter
$\mu$ from the condition that electroweak symmetry is broken radiatively. We
choose the sign of $\mu$ to be negative, since the other choice will give large
$b\rightarrow s\gamma $ rate. The lighter stau mass will be much smaller than
the  other sleptons (even when $tan\beta$ is small) due to the presence of the
new  coupling $f_3$. In figure 3 we show the mass contours of the lighter stau
(dotted line), the $\Delta^{\pm\pm}$ (solid line) and the  lighter
selectron(dot-dashed line) in the
$m_0$ and $tan\beta$ plane for $m_{1/2}=180 GeV$. We choose  
$\tilde\Delta^{\pm\pm}$ mass
$\sim 100$ GeV at the Weak scale (90 GeV at the intermediate scale).  As $m_0$
increases the $\Delta^{\pm\pm}$ mass  decreases due to the subtractive effect in
the RGE originating from the larger soft SUSY breaking scalar mass.   We can see
from the figure that the bound  on $\Delta^{\pm\pm}$ mass rules out the upper
range of $m_0$.
 The lower bound on lighter stau ($\tilde\tau_1$)
 mass on the other hand rules out lower range of
$m_0$.

  The latest bound on lighter stau mass is about
57 GeV \cite{aleph1}.  In this scenario $\tilde\Delta$ mass is too high to be  
pair  produced at LEPII.  We also observe (due to the smallness of $f_{1,2}$
compared to
$f_3$) that the lighter selectron mass is much higher than the  $\tilde\tau_1$
mass even when the
$\tan\beta$ is small for the same $m_0$ and 
$m_{1/2}$ values.  This will differentiate  between the final states of 
chargino pair production in MSSM and in this model.  In fig 4 we show the same 
mass contours for  $\tilde\Delta^{\pm\pm}\sim 80$ GeV at the weak scale (70 GeV
at the intermediate scale). Lower $\Delta^{\pm\pm}$ mass indicates lesser
effect on the 
$\tilde\tau_1$ mass from the new interactions  and consequently more parameter
space for lower
$m_0$, however low
$\Delta$ mass rule out more  parameter space from the higher 
$m_0$ range. In this scenario the $\tilde\Delta$s can be pair produced at LEPII.
In Fig 5 we exhibit a scenario where $\tilde\Delta$ mass is $\sim 120$ GeV at
the
 weak scale (110 GeV at the intermediate scale). The lightest neutralino
($\chi^0_1$)in all these scenarios are $\sim 49-57$ GeV and the  lightest
chargino mass is around 80-90 GeV. If we vary the $m_{1/2}$, e.g 
$m_{1/2}=150 GeV$, the $\chi^0_1$ becomes 33-43 GeV and the lightest  chargino
becomes 57-67 GeV. the scalar masses also get reduced by 10 -20 GeV. 
 In Fig 6 we show a scenario where the third generation coupling is smaller
$f_3\sim 0.25$. As one can expect, the effect of the new couplings in the 
stau
mass is reduced. For low $tan\beta$, as in the conventional SUGRA mode, the
$\tilde\tau_1$ and the lighter selectron mass ($\tilde e_R$)  become very
close. The  $\Delta$ boson mass lower than 88 GeV appears for 
$m_0\rangle 270 $ GeV, where as in Fig. 3 the same mass contour appears at
$m_0\rangle 140 $ GeV (In both the cases the $m_{1/2}$ at the GUT scale and
$M_{\tilde\Delta}$ at the intermediate scale are same. From the analysis of the
parameter space we surmise that most of the allowed parameter space  for large
$f_3$ can be searched in the present colliders.
\begin{figure}[htb]
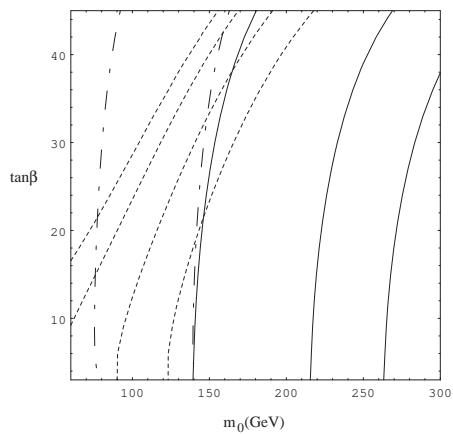

\vspace{0 cm}
\centerline{ \DESepsf(dmdeltafig3.epsf width 6 cm) }
\smallskip
\caption{The mass contours for $M_{\tilde\Delta^{\pm\pm}}=90$GeV at the 
intermediate scale and $f_3=0.5$  are shown. The $\tilde\tau_1$ mass contours
(dashed line) from left to right   depict 45, 60, 80 and 100 GeV masses, the
$\Delta^{\pm\pm}$ mass contours (solid lines)  from left to right depict 88,70
and 50 GeV masses, the $\tilde e_R$ ($\tilde e_R$ and $\tilde\mu_R$ masses are
same)  mass contours (dot dashed line) from left to right depict  100 and 140 GeV
masses}
\end{figure}

\begin{figure}[htb]
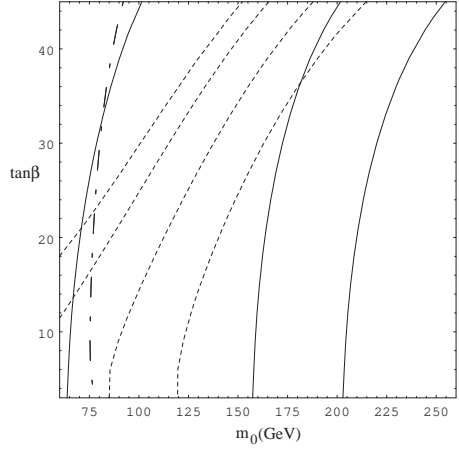

\vspace{0 cm}
\centerline{ \DESepsf(dmdeltafig4.epsf width 6 cm) }
\smallskip
\caption{The mass contours for $M_{\tilde\Delta^{\pm\pm}}=70$GeV at the 
intermediate scale and $f_3=0.5$  are shown. The $\tilde\tau_1$ mass contours
(dashed line) from left to right depict  45, 60, 80 and 100 GeV masses, the
$\Delta^{\pm\pm}$ mass contours (solid lines)  from left to right depict 80, 65
and 50 GeV masses, the $\tilde e_R$  mass contour (dot dashed line) depicts  100
GeV mass}
\end{figure}
\begin{figure}[htb]
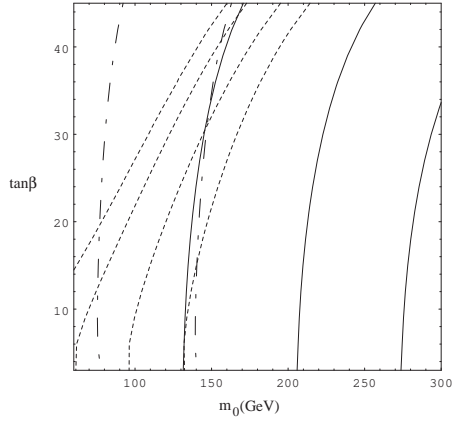

\vspace{0 cm}
\centerline{ \DESepsf(dmdeltafig5.epsf width 6 cm) }
\smallskip
\caption{The mass contours for $M_{\tilde\Delta^{\pm\pm}}=110$GeV at the 
intermediate scale and $f_3=0.5$  are shown. The $\tilde\tau_1$ mass contours
(dashed line) from left to right depict  45, 60, 80 and 100 GeV masses, the
$\Delta^{\pm\pm}$ mass contours (solid lines)  from left to right depict  108, 95
and 70 GeV masses, the $\tilde e_R$  mass contours (dot dashed line) from left to
right depict  100 and 140 GeV masses}
\end{figure}
\begin{figure}[htb]
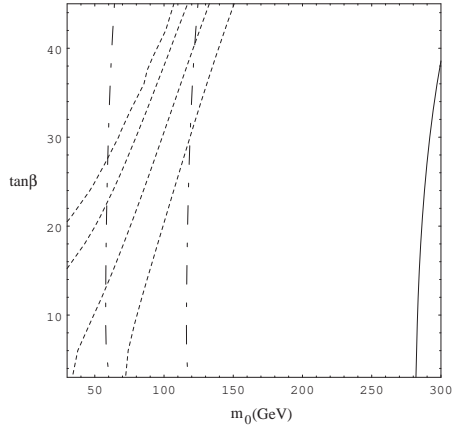

\vspace{0 cm}
\centerline{ \DESepsf(dmdeltafig6.epsf width 6 cm) }
\smallskip
\caption{The mass contours for $M_{\tilde\Delta^{\pm\pm}}=90$GeV at the 
intermediate scale and $f_3=0.25$  are shown. The $\tilde\tau_1$ mass contours
(dashed line) from left to right depict  45, 60, 80 and 100 GeV masses, the
$\Delta^{\pm\pm}$ mass contour (solid lines)  depicts 88 GeV, the $\tilde e_R$
mass contours (dot dashed line) from left to right depict  100 and 140 GeV
masses. }
\end{figure}

The $\Delta$ will decay into a pair of $\tau$s (since the coupling to the 
other
leptons are suppressed). The $\tau$'s have high $p_T$. We have 4 $\tau$s 
in the
final state. The like charged $\tau$s originate
from the same vertex. 

The SM background for this process can come from the 
pair production of $Z^0$ and the  subsequent decay of each $Z^0$ to a
$\tau^{+}\tau^{-}$ pair (in all these cases  oppositely charged $\tau$s come
from the same vertex). This event will  have no missing energy, but the rate is
small due to the small branching ratio  for
$Z^0\rightarrow\tau^{+}\tau^{-}$ ($\sigma.B^2\sim 10^{-3} pb$). The $4\tau$ 
background  can also come from the production of $Z\gamma^*$, 
$\gamma^*$  converting into $\tau^{+}\tau^{-}$ and the $Z$ decaying into 
$\tau^{+}\tau^{-}$ pair.  Another source can be the production of two  virtual
photons along with $e^{+}e^{-}$ and each  photon converting into 
$\tau^{+}\tau^{-}$. However, in both the above processes, the cross section is
$\sim 10^{-3}$pb. Thus, the SM background is negligible small for the 4$\tau$
signal. 
 
The $\tilde \Delta$ will primarily decay into $\tilde\tau_1$ and  $\tau$
(almost 100$\%$).  The $\tilde\tau_1$ will then decay into  $\tau$ and
$\chi^0_1$ (missing energy) (100$\%)$. The final state has 4
$\tau$'s plus missing energy. Two of the $\tau$s have high
$p_T$ and these originate from the decay of lighter stau.  This kind of 4
$\tau $ plus missing energy signal also originate from the $\chi^0_1$ pair
production in the GMSB scenarios where  lighter stau is the NLSP. But there is a
subtle difference in the final state which we  will discuss in the next
section.  

The chargino pair production can also give rise to
$2\tau$ plus missing energy states. Since the staus are much lighter than 
the other
sleptons,  the chargino will primarily decay into
$\tilde\tau_1$ and  $\nu_{\tau}$ in the leptonic decay channel.  On the 
other hand,
 the  chargino decays into
$e$'s and $\mu$'s in MSSM. The charginos can be pair  produced at LEP II and
Tevatron. The production of chargino and the second lightest neutralino (this
crosssection is larger than the chargino pair production ) at the
Tevatron will also give rise to lots of high
$p_T$ taus in the final state. The second lightest neutralino ($\chi^0_2$)
 primarily decays into  
$\tau$ and $\tilde\tau_1$ and $\tilde\tau_1$ then decays into a tau
and the  lightest neutralino. Altogether, there can be 3$\tau$s along
with missing energy. Since $\chi^0_2$ mass is much larger than the 
$\tilde\tau_1$ mass, all the three $\tau$s will have high $p_T$.

The $\Delta$ bosons and the $\Delta$-inos can be pair produced at the LEPII  and
at the Tevatron as well. The productions of $\Delta$'s (fermion and boson) at 
LEP II and $\Delta$ scalar at Tevatron have been
considered in the references\cite{huitu}.  We give the production 
cross-sections of the $\Delta $ 
bosons at the LEP II and at the Tevatron in Figures 7 and 8. The
$\Delta$-ino is  pair produced at the LEP II and at the  Tevatron via Z,
$\gamma$ exchange. Usually there is also a selectron mediated t-channel 
contribution in the case of $\Delta$-ino production at the  
$e^{+}e^{-}$ collider. However the contribution from this diagram in this
 model is negligible since the $\Delta$ coupling to the 
first generation leptons is very small. 
The production crosssections are larger than the scalar
counterparts for the same mass. 
 In Figure 9  we show the production
 of $\tilde\Delta$ at LEPII for $\sqrt s=182 $ GeV and 194 GeV. We can see  that
the $\Delta$-ino mass of  95 GeV gives 
rise to a cross section of  3 pb at LEPII.
Thus, if the  $\Delta$-inos are produced at LEPII  the cross section will be 
quite large. 

In Figure 10 we show the production cross-section at the Tevatron for
$\sqrt s=2 TeV.$  The production crosssection is about 0.7 pb at $\sqrt s=2$TeV 
for $\Delta$-ino mass of 95 GeV and the cross section is about 0.6 pb at 
$\sqrt s=1.8$TeV for the  same $\Delta$-ino mass. Hence  with 110$pb^{-1}$ of
already accumulated luminosity, the number of events are $\sim 66$. Since the
final state are pure $\tau$ leptons, detection is difficult.
 But we will have some events left even after taking the $\tau$ detection
efficiency to be
 small.  We urge the experimentalists to look for the $\tau $ signals  in the
data which has been already accumulated and also in the data that will be
 generated in the future runs.
\begin{figure}[htb]
\vspace{0 cm}
\centerline{ \DESepsf(dmdeltanew.epsf width 6 cm) }
\smallskip
\caption{The $\Delta^{\pm\pm}$ pair production GeV 
at the cross section at LEPII. The
solid and the dotted line correspond to the cross sections at the center of mass
energies to be 194 GeV and 182 GeV.}
\end{figure} 

\begin{figure}[htb]
\vspace{0 cm}
\centerline{ \DESepsf(dmdeltanew1.epsf width 7 cm) }
\smallskip
\caption{The $\Delta^{\pm\pm}$ pair production cross section at 
 Tevatron (center of mass energy is 2 TeV)}
\end{figure}
\begin{figure}[htb]
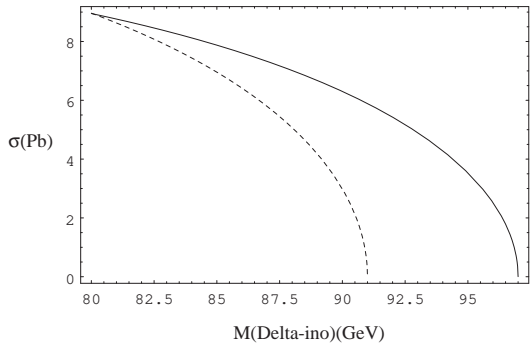

\vspace{0 cm}
\centerline{ \DESepsf(dmdeltafig7.epsf width 7 cm) }
\smallskip
\caption{The $\tilde\Delta^{\pm\pm}$ pair production cross section at LEP II. 
The solid line and the dotted line correspond to the cross section at the
  center
of mass energies to be 194 GeV and 182 GeV respectively.}
\end{figure}

\begin{figure}[htb]
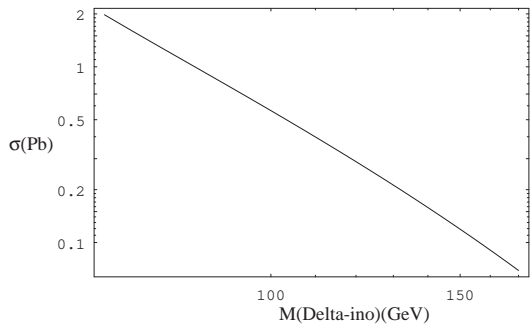

\vspace{0 cm}
\centerline{ \DESepsf(dmdeltafig8.epsf width 7 cm) }
\smallskip
\caption{The $\tilde\Delta^{\pm\pm}$ pair production cross section at  Tevatron
(center of mass energy is 2 TeV)}
\end{figure}

\section{Gauge mediated susy breaking scenario}

In the previous section we assumed that the scale at which  supersymmetry is
broken is higher than the $W_R$ scale. However this need not be the case and in
particular there has recently been a lot of interest in theories where gauge
interactions are the mediators  of supersymmetry breaking at a relatively low
scale\cite{dine}.  Let us discuss the implications of this scenario for our
model. The soft SUSY breaking terms are now generated explicitly  only at  the
scale at which the messenger fields are integrated out, and are not explicitly
present at the $W_R$ scale. Since they are generated by  loop graphs involving
the gauge bosons of the residual symmetries, their form will be such as to
respect only the surviving gauge symmetries.
 We will show that this difference  has consequences for phenomenology. 
 Let us assume for simplicity that the
messenger sector consists of a vectorlike isosinglet pair of fields  (charge
-1/3)
$Q\oplus \bar{Q}$ and a vectorlike weak isodoublet pair $L\oplus \bar{L}$.

As shown in \cite{chacko}, in this case, the scalar $\Delta$ mass has two
contributions at the SUSY breaking scale of $100$ TeV or so: one coming from  
the two loop gauge contributions in the usual manner (for a  review see
\cite{martin}) and another coming from the higher dimensional terms. On the
other hand, the $\tilde{\Delta}$ mass gets contribution  only from the latter
kind of terms. We therefore expect that in the gauge mediated scenario, the
$\tilde{\Delta}$ will be lighter of the  two particles.  

The soft SUSY breaking gaugino and the scalar masses at the messenger scale $M$
are given by \cite{martin}
\begin{center}
$\tilde M_i(M) = n\,g\left({\Lambda\over M}\right)\,
{\alpha_i(M)\over4\pi}\,\Lambda$
\end{center} and
\begin{center}
$\tilde m^2(M) = 2 \,n\, f\left({\Lambda\over M}\right)\,
\sum_{i=1}^3\, k_i \, C_i\,
\biggl({\alpha_i(M)\over 4\pi}\biggr)^2\,
\Lambda^2 $
\end{center}  where $\alpha_i\; (i=1-3)$ are the three SM gauge couplings and
$k_i=1,1,3/5$ for SU(3), SU(2), and U(1), respectively. The $C_i$ are zero for
gauge singlets, and 4/3, 3/4, and $(Y/2)^2$ for the fundamental representations
of $SU(3)$ and $SU(2)$ and $U(1)_Y$ respectively  (with $Y$ defined by
$Q=I_3+Y/2)$. Here $n$ corresponds to
$n((Q,L)+{\bar {(Q,L)}})$. $g(x)$ and $f(x)$ are messenger scale threshold
functions  with
$x=\Lambda/M$. 

We calculate the SUSY mass spectrum using the appropriate RGE equations
\cite{barger} with the boundary conditions given by the equations above  and we
vary
 five free parameters $\Lambda$, $M/\Lambda$, $tan\beta$, $n$ and the sign of
$\mu$ ($\mu$ is the coefficient of the bilinear Higgs term in the
superpotential). We first run the Yukawa couplings (along with the three new 
couplings $f_{1,2,3}$) and gauge couplings from the weak scale upto the GMSB 
scale. At the GMSB scale we use the boundary conditions and then use the
necessary  RGEs for the soft SUSY breaking masses in order  run down to the weak
scale.

 The CLEO constraint on the  $b\rightarrow s\gamma $ rate  restricts $\mu<0$
\cite{dwt}.  In the absence of late inflation, cosmological constraints put an
upper bound on the gravitino mass of about $10^4$ eV
\cite{pp}, which restricts $M/\Lambda=1.1-10^4$. In the figures we show the 
results for  n=1, but we discuss the other values of n also. For reasons 
discussed before, we assume that $f_{1,2}$ are small ( $\sim 0.05$), but 
$f_3$ 
is larger. We show our result for $f_3\sim 0.5$ and 
$f_3\sim 0.25$. We also vary  $M_{\tilde\Delta}$ between 70-120 GeV at the 
GMSB scale.

The gravitino is always the LSP. In the usual GMSB case the $\chi^0_1$ and the
$\tilde\tau_1$ fight for the NLSP spot. In this model the
$\tilde\Delta^{\pm\pm}$ also joins the  race to become NLSP. The third
generation lighter stau mass gets affected due to the presence of the additional
large coupling
$f_{3}$. Thus  the $\tilde\tau_1$ is much smaller compared to the conventional
GMSB case for the same  parameter space. Consequently the lighter stau will be
the NLSP for a wider region of parameter space compared to the lighter
neutralino.  

In figure 11a we have shown  the mass contours of $\tilde\tau_1$
(solid line),
$\tilde e_R$ (dotted line), $\chi^0_1$  (dash-dotted line) and the chargino
masses (dashed line) for
$M=1.1 \Lambda$, n=1 and $M_{\tilde\Delta}=90$ GeV at the GMSB scale (94 GeV 
at the weak scale). We also  show the contour along which the lighter stau mass
and the neutralino mass are same (thick solid  line). The region above the
contour has lighter stau as the NLSP. We see that only a small region for 
$tan\beta$ 3-15 and 
$\Lambda$ 40-60 TeV has $\chi^0_1$ as NLSP. When $\chi^0_1$ is the
NLSP, it decays into a photon and a gravitino. If $\chi^0_1$ is pair produced at
LEPII, the final state has $\gamma\gamma$ plus
${\rlap/E}_T$. The photons are  hard and easy to detect. Already we have  bound
on the $\chi^0_1$  mass of around 80 GeV at LEP II \cite{aleph2} 
provided the selectrons are not too heavy. We can see
that if we use $\chi^0_1$ mass bound as 80 GeV, the region left out in fig 11a
where
$\chi^0_1$ is the NLSP is very small. The region where $\tilde\tau_1$ is the
NLSP, which is the dominant region, stau decays into a high $p_T$ $\tau$ and a
gravitino (missing energy). So far there is not much bound in these regions,
other than the $\tilde\tau_1$ has to be larger than 57 GeV. In the figures we
have shown the stau mass contours of 45, 60, 80 and 100 GeV. The
$\tilde\tau_1$ mass contours have large  dependence  on the
$tan\beta$, the $\tilde\tau_1$ decreases with the increase in $\tan\beta$.  
The $\tilde e_R$, chargino and the $\chi^0_1$ mass contours do not have much
$tan\beta $ dependence.  

In figure 11b  we show the same mass contours in the
plane of $\Lambda$ and $M/\Lambda$ for
$\tan\beta=10$. The contours along which $\tilde\tau_1$ mass is equal to the
$\chi^0_1$ mass form an envelope. Within the envelope $\chi^0_1$ is NLSP. The
$\tilde e_R$  masses increase as the $M/\Lambda$ ratio increase. The couplings
$f_{1,2}$ are small to  affect the $\tilde e_R$ or the $\tilde \mu_R$ masses
(selectron  and smuon masses are almost same). The lighter stau mass however
decreases with the increase in the ratio of $M/\Lambda$ (subtractive effect
from the soft scalar  masses due to the presence of the new coupling overcome
effects coming from  the gaugino masses in the RGE). In the conventional GMSB
model the stau mass increases  with the increase in the $M/\Lambda$ ratio. 
With the  improvement of the
$\tilde\tau_1$ mass bound  much more parameter space will be ruled out in the
higher$M/\Lambda$ ratio. The $\chi^0_1$ and  the chargino masses initially
decrease with the increase in the ratio of $M/\Lambda$ due to the threshold
corrections. The 80 GeV $\chi^0_1$ bound rules out much of the parameter
space in the envelope where$\chi^0_1$ is the NLSP. The
$M_{\tilde\Delta^{\pm\pm}}$ varies  between 94-108 GeV at the weak scale.
Unlike the SUGRA scenario, the $\Delta^{\pm\pm}$ mass is much larger than the
fermionic part at the weak scale and its mass, in the full range shown in the
figure, varies from 170 -570 GeV (due to the large soft SUSY breaking
contribution at the GMSB scale). Thus, the allowed parameter space in GMSB
scenarios are much more than the SUGRA scenarios. 

In figures 12 and 13, we have shown the mass contours in the
$\Lambda-M/\Lambda$ plane for $M_{\tilde\Delta^{\pm\pm}}=70$  and 110 GeV at the
GMSB scale. The $tan\beta$ is chosen to be 10. When 
$M_{\tilde\Delta^{\pm\pm}}=70$ GeV, the envelope  is larger, since
$\tilde{\tau_1}$ mass has less subtractive contribution from the delta mass. On
the other hand when
$M_{\tilde\Delta^{\pm\pm}}=110$ GeV, the envelope shrinks. The envelope can
also  increase in the size if we have smaller coupling $f_3$. We show the
effect of a smaller
$f_3$ in figure 14. This figure looks more like the conventional GMSB model. We
see that there is no parameter space  where $\tilde{\tau_1}$ is NLSP. But if we
increase
$tan\beta$ we will hit the  region where $\tilde{\tau_1}$ is the NLSP. The
$\tilde{\tau_1}$ mass in this case, as expected,  increases with the increases
in the ratio of
$M/\Lambda$. So far in all these figures we have used n=1. The effect  of larger
values of n can easily be surmised from the mass formula. As n increases gaugino
masses increase proportionally, on the other hand the scalar masses increase as
$\sqrt n$. Hence stau mass is the NLSP for a even wider region of parameter
space. The lighter selectron's mass becomes closer to the lighter stau mass and 
lower than the neutralino mass. 

Let us now discuss the signals. At LEP II, the main production processes  are 
the $\chi^0_1$ pair, the $\tilde\Delta^{\pm\pm}$ pair, the $\tilde e_R$ 
pair and the $\tilde{\tau_1}$ pair. 

In the case of $\tilde\Delta^{\pm\pm}$  pair production, each
$\tilde\Delta^{\pm\pm}$ wll decay primarily into a $\tilde{\tau_1}$ and a tau
(both having same sign of charge). The other decay modes involving e.g. a
electron and a 
$\tilde e_R$ and a muon and a smuon are suppressed primarily because of small
$f_{1,2}$ couplings. In the parameter space when stau is the NLSP,
 $\tilde\Delta^{\pm\pm}$ decays into a $\tilde\tau_1$ and a $\tau$ and 
$\tilde\tau_1\rightarrow\tau\tilde G$. Thus, the final
state has  4$\tau$ plus missing
energy and out of the 4
$\tau$s, two   have high
$p_T$ (higher $p_T$ than the SUGRA case). These two high $p_T$ $\tau$s have
opposite sign electric charges. 

The  $\chi^0_1$ pair production also gives rise to 4
$\tau$s plus missing energy (since each $\chi^0_1\rightarrow\tilde\tau_1\tau$),
with two of these
$\tau$s having high
$p_T$
\cite{ddn}. However there is  an essential difference between the signal in this
case and in the previous case. Since the neutralino is a majorana particle, the
two high 
$p_T$ $\tau$s can have same or opposite sign (with equal probability) for 
the electric charges and thus providing a way to discrminate between the two
cases.

 In the case of 
$\tilde e_R$ pair productions, each selectron will either decay into a electron
and a $\chi^0_1$  or a $\tilde\Delta^{\pm\pm}$ and a positron (if the selectron
mass is higher than  the $\chi^0_1$ mass and the $\tilde\Delta^{\pm\pm}$ mass).
Both $\chi^0_1$ and $\tilde\Delta^{\pm\pm}$ decay to $2\tau$ plus missing
energy. The final states in either case will have $2e 4\tau$ plus missing energy
with  two of the
$\tau$s having high 
$p_T$. Their relative sign will determine the decay channel of the 
selectron. If, however, the $\tilde e_R$ mass is lower than both the
$\chi^0_1$ mass  (large n case) and the $\tilde\Delta^{\pm\pm}$ mass,  the
selectron can decay into an  electron and a gravitino or via offshell production
of the
$\chi^0_1$ or 
$\tilde\Delta^{\pm\pm}$. The $\chi^0_1$ and
$\tilde\Delta^{\pm\pm}$ then convert into $2\tau$ plus missing energy. In the
former case the final state of the selectron pair production has $2 e$ plus
missing energy and in the later case the signal is $2e4\tau$ plus missing enrgy.
Depending on the  parameter space, these offshell decay modes can be comparable
or greater than the onshell decay mode \cite{amK}.  In the case of smuons the
electrons in the final states will be replaced by the muons.

When $\chi^0_1$ is the NLSP, the
$\tilde\Delta^{\pm\pm}$ will  decay into a $\tilde\tau_1$ and  a $\tau$
(100$\%$). The
$\tilde\tau_1$ will decay into a $\chi^0_1$ and a 
$\tau$. The $\chi^0_1$ then decays into a photon and a gravitino. All the 
branching ratios are  100 $\%$. The final state from the 
$\tilde\Delta^{\pm\pm}$ pair production will have $2\gamma 4\tau$ plus missing 
energy-which is a spectacular signal and very hard to miss. This signal will
appear along with the electrons or the muons in the case of selectron or smuon
production, where the selectron decays as described above i.e. $\tilde
e_R\rightarrow e\tilde\Delta^{\pm\pm}$. 

In the conventional GMSB models when
$\chi^0_1$ is the NLSP case, one gets  at most 2 leptons along with 2
$\gamma$ (through the slepton productions whose cross section is much smaller
than the delta-ino production cross section) at the LEP II and at the Tevatron or
3  leptons plus 2 photons (chargino-second lightest neutralino production) 
at the Tevatron. Thus the signal
$2\gamma 4\tau$plus missing  energy will clearly distinguish this model from the
ordinary GMSB models in the parameter space where $\chi^0_1$ is the NLSP. It
may also happen that the $\tilde\Delta^{\pm\pm}$ mass is smaller than the
$\tilde\tau_1$ or the
$\chi^0_1$ mass. In that case the $\tilde\Delta^{\pm\pm}$ will decay into a
$\tau$ and a virtual $\tilde\tau_1$ which will convert into a $\tau$ plus
gravitino or if the $\chi^0_1$ mass is lower than the $\tilde\tau_1$ mass, then
the
$\tilde\tau_1$ will decay into a
$\tau$ and a $\chi^0_1$. The $\chi^0_1$ will then convert into a photon and the
gravitino. Thus the final states are same as discussed in the cases when
$\chi^0_1$ is the NLSP or $\tilde\tau_1$ is the NLSP.
\begin{figure}[htb]
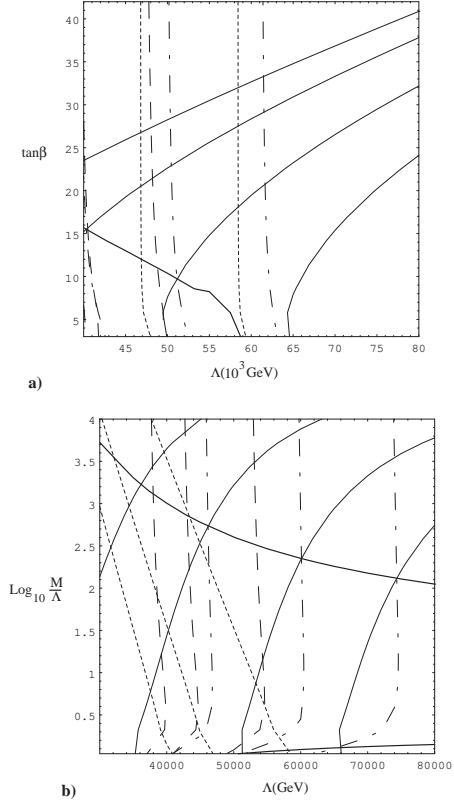

\vspace{0 cm}
\centerline{ \DESepsf(dmdeltafig9ab.epsf width 6 cm) }
\smallskip
\caption{a) The mass contours for $M_{\tilde\Delta^{\pm\pm}}=90$GeV at the  GMSB
scale and $f_3=0.5$  are shown. The $\tilde\tau_1$ mass contours (solid line)
from left to right depict  45, 60, 80 and 100 GeV masses, the $\tilde e_R$  mass
contours (dotted lines)  depict 80 (along the left axis), 95 and 115 GeV
masses, the
$\chi^0_1$ (dot-dashed) mass contours depict 60, 80 and 100 GeV masses, the
lightest chargino ($\chi^{\pm}$) mass contours  (dashed) depict  94, 100 and
130 GeV masses b) The mass contours for
$M_{\tilde\Delta^{\pm\pm}}=90$GeV at the  GMSB scale and $f_3=0.5$  are shown.
The $\tilde\tau_1$ mass contours (solid line) from left to right depict  45, 60,
80 and 100 GeV masses, the $\tilde e_R$  mass contours (dotted lines)
 depict 80,
95 and 115 GeV masses, the $\chi^0_1$ (dot-dashed) mass contours  depict 60, 80
and 100 GeV masses, the lightest chargino  mass contours (dashed) depict 
 88, 100 and 130 GeV masses. The thick solid lines in both
the figures depict the contour along which the  lighter stau mass equals to the
lightest neutralino mass.}
\end{figure}
\begin{figure}[htb]
\vspace{0 cm}
\centerline{ \DESepsf(dmdeltafig10.epsf width 6 cm) }
\smallskip
\caption{The mass contours for $M_{\tilde\Delta^{\pm\pm}}=70$GeV at the  GMSB
scale and $f_3=0.5$  are shown. The $\tilde\tau_1$ mass contours (solid line)
from left to right depict 
 45, 60, 80 and 100 GeV masses, the $\tilde e_R$  mass contours (dotted lines) 
depict  75, 95 and 115 GeV masses, the $\chi^0_1$ (dot-dashed) mass contours 
depict 60, 80 and 100 GeV masses, the lightest chargino  mass contours
(dashed)depict  70, 100 and 130 GeV masses. The thick solid lines depict the
contour along which the  lighter stau mass equals to the lightest neutralino
mass.}
\end{figure}
\begin{figure}[htb]
\vspace{0 cm}
\centerline{ \DESepsf(dmdeltafig11.epsf width 6 cm) }
\smallskip
\caption{The mass contours for $M_{\tilde\Delta^{\pm\pm}}=110$GeV at the  GMSB
scale and $f_3=0.5$  are shown. The $\tilde\tau_1$ mass contours (solid line)
from left to right depict 
 45, 60, 80 and 100 GeV masses, the $\tilde e_R$  mass contours (dotted lines) 
depict 95 and 115 GeV masses, the $\chi^0_1$ (dot-dashed) mass contours  depict
60, 80 and 100 GeV masses, the lightest chargino  mass contours( dashed)depict 
100 and 130 GeV masses. 
The thick solid lines depict the contour along which the 
lighter stau mass equals to the lightest neutralino mass.}
\end{figure}

\begin{figure}[htb]
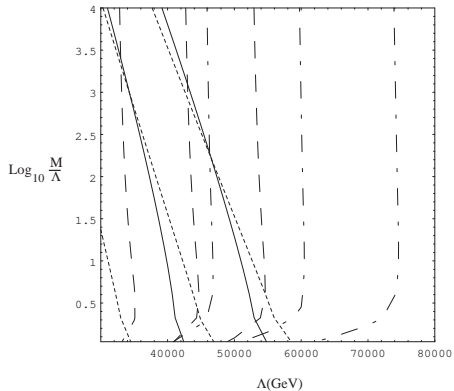

\vspace{0 cm}
\centerline{ \DESepsf(dmdeltafig12.epsf width 6 cm) }
\smallskip
\caption{The mass contours for $M_{\tilde\Delta^{\pm\pm}}=90$GeV at the  GMSB
scale and $f_3=0.25$  are shown. The $\tilde\tau_1$ mass contours (solid line)
from left to right are 
 80 and 100 GeV, the $\tilde e_R$  mass contours (dotted lines)  are 75, 95 and
115 GeV, the $\chi^0_1$ (dot-dashed) mass contours  are 60, 80 and 100 GeV, the
lightest chargino  mass contours( dashed)are  70, 100 and 130 GeV. }
\end{figure}  
\section{conclusion}

To summarize, we have studied the phenomenological implications and the
collider signatures of the remnant doubly charged Higgs boson 
$\Delta^{\pm\pm}$ and its fermionic partner the Deltaino 
$\tilde\Delta^{\pm\pm}$. We find that existing limits on lepton flavor 
violation and  muonium-anti-muonium transition can be used to conclude that the
dominant coupling of these particles are diagonal and mostly to the third
generation of leptons. They of course are forbidden from coupling to the quarks
by electric charge conservation. The effect of the dominant third generation
lepton coupling has the consequence that  the $\tilde{\tau_1}$ mass is much
smaller than the other slepton masses. This  gives rise to mulitiple $\tau$
enriched signal at LEP II and Tevatron.  The fermionic as well as the bosonic
partner of the doubly charged Higgs bosons can be produced at the present
colliders and the signals contain 4 $\tau$ (plus 2$\gamma$ in some scenarios of
the GMSB version of the models) with and without missing energy. This signal is
found to be detectable  for reasonable range  of mass values of the particles
and could be used to test the supersymmetric left-right models of the type
discussed here or to restrict the allowed parameter range of the model. Most of
the  allowed parameter space can be searched in the existing colliders.  We
therefore urge the experimentalists to analyze the tau events in the existing 
as well as in the future data.

\noindent{\bf Acknowledgement}

Work of R. N. M. has been supported by the National Science Foundation grant
No. PHY-9421385 and that of B. D. by DE-FG06-854ER 40224. We thank  G.
Altarelli, J. Gunion and G. Snow for encouragement and comments.

\section{Appendix}   The
standard MSSM RGEs  for the Yukawa couplings and the soft SUSY breaking terms
will be modified due to the presence of the new couplings and the fields,. 
There will be new  RGEs for the additional fields and the couplings. The new
interaction involves the  third generaion righthanded leptons and the new field
$\Delta^{\pm\pm}$.  We will keep only the $f_3$ coupling in the RGE's  (in
our numerical calculation we use all of them). The new RGEs and the  modified
ones are listed below:
\begin{eqnarray} 2{\cal D}\lambda_{\tau} &=&\lambda_{\tau}(-
\sum_i{C\tau_i(4\pi\alpha_i)}+
\lambda_{b}^2+4 \lambda_{\tau}^2+4 {f_3}^2),
\end{eqnarray} where $C\tau_i=9/5,3,0$ and 
${\cal D}\equiv{{16\pi^2}\over 2}{d\over{dt}}$.

\begin{eqnarray} 2{\cal D}f_3 &=&f_3(-
\sum_i{Cf_{3i}(4\pi\alpha_i)} +10 {f_3}^2+4 \lambda_{\tau}^2),
\end{eqnarray} where $Cf_{3i}=36/5,0,0$.

\begin{eqnarray} {\cal D}m_{\tau^c}^2 &=& -{48\over 5}\pi\alpha_1{\tilde M_1}^2
+ 2\lambda_{\tau}^2(m_{\tau_c}^2+m_{\tau}^2+M_{H1}^2+A_{\tau}^2)\\\nonumber &+&
4 {f_3}^2(m_{\Delta^{\pm\pm}}^2+M_{\tau^c}^2+{A_{3}}^2),
\end{eqnarray}  
\begin{eqnarray} {\cal D}m_{\Delta^{\pm\pm}}^2 &=& -{192\over
5}\pi\alpha_1{\tilde M_1}^2+ 2 {f_3}^2(m_{\Delta^{\pm\pm}}^2+2
m_{\tau^c}^2+{A_{3}}^2),
\end{eqnarray}
  We do not include the effects of the couplings to the other generation 
$f_{1,2}$ since they are very small.
\begin{eqnarray} {\cal D}A_{\tau} &=&(\sum_i{C\tau_i4\pi\alpha_i M_i} +
\lambda_{b}^2 A_b+4 \lambda_{\tau}^2 A_{\tau}+4 {f_3}^2 A_3),
\end{eqnarray} 

\begin{eqnarray} {\cal D}A_3 &=&( (4\pi\alpha_1){36\over 5} M_1 +10 {f_3}^2
A_3+4
\lambda_{\tau}^2 A_{\tau}),
\end{eqnarray} 

\begin{eqnarray} 2{\cal D}M_{\tilde\Delta^{\pm\pm}} &=&M_{\tilde\Delta^{\pm\pm}}
( {-(4\pi\alpha_1){48\over 5} M_1} +4 {f_3}^2).
\end{eqnarray}

\end{document}